\documentclass[
  aps,
  prb,
  reprint,
  amsmath,
  raggedbottom,
  superscriptaddress,
  longbibliography
]{revtex4-2}

\usepackage{graphicx}
\usepackage{amssymb}
\usepackage{bm}
\usepackage{enumitem}

\usepackage[svgnames]{xcolor}
\usepackage{hyperref}
\hypersetup{
    pdfstartview=FitH,
    pdfpagemode=UseOutlines,
    colorlinks,
    citecolor=blue,
    linkcolor=blue,
    urlcolor=blue
}

\begin{document}

\raggedbottom

\title{Breaking the Wiedemann--Franz limit in thermoelectrics via separated flat bands}

\author{Illia Serhiienko}
\email{illia.serhiienko@tuwien.ac.at}
\affiliation{Institute of Solid State Physics, TU Wien, 1040 Vienna, Austria}
\author{Fabian Garmroudi}
\affiliation{Materials Physics Applications—Quantum, Los Alamos National Laboratory, Los Alamos, New Mexico 87545, USA}
\author{Stefan Enzner}
\affiliation{Universität Würzburg, Germany}
\author{Izem Dural}
\affiliation{Universität Würzburg, Germany}
\affiliation{Laboratoire des Solides Irradiés, CEA/DRF/IRAMIS, CNRS, École Polytechnique, Institut Polytechnique de Paris, F-91128 Palaiseau, France}
\author{Giorgio Sangiovanni}
\affiliation{Universität Würzburg, Germany}
\author{Andrej Pustogow}
\email{pustogow@ifp.tuwien.ac.at}
\affiliation{Institute of Solid State Physics, TU Wien, 1040 Vienna, Austria}


\begin{abstract}
Controlling the charge and spin degrees of freedom of electrons in solids has been at the heart of condensed-matter science for centuries. One hallmark of metallic conduction is the Wiedemann-Franz law which implies that a moving charge carries entropy. Despite centuries of dedicated research, disentangling charge and heat transport has remained an unsolved issue so far.
Here we present a direct route to reduce thermal conductivity of conduction electrons with respect to their electrical conductivity via energy-dependent scattering. By constraining electronic transport to a boxcar-type distribution asymmetrically around the Fermi energy, a large Seebeck coefficient and electrical conductivity can be realized simultaneously while electronic heat conduction is suppressed. Based on the case of monolayer Ni$_3$In, which comprises two flat bands around the Fermi energy, we propose flat-band systems as a promising, tunable platform for scattering phase space engineering.
Besides this novel approach, our large-scale assessment of $\approx 5\times 10^4$ data sets in a 'Wiedemann-Franz plot' of $zT$ vs. $S^2$ provides an effective tool to identify reports of $L\ll L_0$ from literature -- pointing towards either new and interesting physical mechanisms -- or overlooked measurement artifacts.
\end{abstract}
\maketitle
Electrons are elementary particles that carry quanta of charge and entropy. Hence, moving an electron brings about an electrical current as well as a transport of entropy, the latter yielding a heat flow. Unlike in vacuum, electrons in solids do not propagate completely independently but can instead be described as quasiparticles interacting with each other and the crystal lattice. This way, electronic degrees of freedom may be disentangled from each other. A well-known example is spin-charge separation that has been intensely studied in one-dimensional Luttinger liquids \cite{jompol_probing_2009} and, more recently, also in quantum spin liquids \cite{Balents2010,savary_quantum_2017}.
While such extreme behavior is typically found in correlated electron systems nearby electronic instabilities, for instance around the van Hove singularity of Sr$_2$RuO$_4$~\cite{Stangier2022}, common metals usually follow the widely established Wiedemann-Franz (WF) law $\kappa_{\rm e}=L\sigma T$, which relates the electronic thermal conductivity $\kappa_{\rm e}$ to the electrical conductivity $\sigma$ via the Lorenz number $L$. The recent advent of \textit{metallic thermoelectrics}~\cite{Garmroudi2023,garmroudi_topological_2025,garmroudi_energy_2025,Iwasaki2026}, where the figure of merit simplifies in good approximation to $zT\approx S^2/L$, calls more than ever for materials with Lorenz numbers reduced below the Sommerfeld value $L_{0}$ in order to boost efficiency~\cite{garmroudi_energy_2025}.

In this work, we establish a general framework for reducing the Lorenz number in metallic materials by 
energy-dependent interband scattering. First, we theoretically show and revisit that a boxcar-type transport distribution is required to suppress the Lorenz number well below $L_{0}$ while maintaining a large $S$ and $\sigma$ 
-- in agreement with earlier considerations. We then show that a pseudo-boxcar transport function can be realized through flat-band-induced energy-selective scattering, where two or more flat bands separated in energy generate the two scattering edges. 
Finally, we outline potential material platforms, such as monolayer Ni$_3$In, that could experimentally realize such a scenario.

\section{State of the Art}

\begin{figure}[b]
 \centering
 \includegraphics[width=1.00\columnwidth]{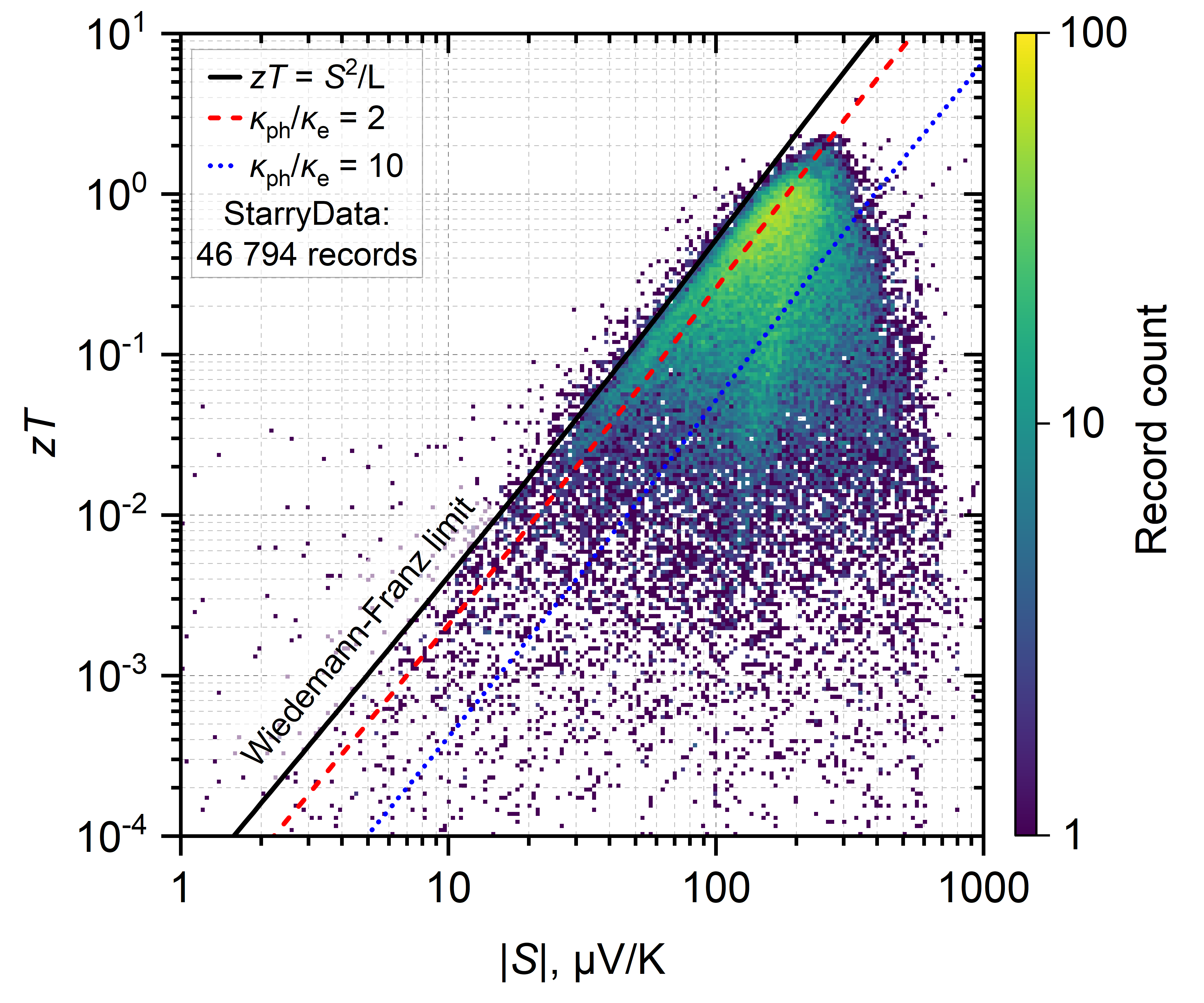}
 \caption{ Wiedemann-Franz plot of $zT$ versus $|S|$~\cite{Garmroudi2023} for 46,794 thermoelectric records from the StarryData database~\cite{katsura_starrydata_2025} in the 100--1000\,K range. The solid line $zT = S^{2}/L_{0}$ marks the Wiedemann--Franz (WF) limit, approached for $\kappa_{\rm ph}\ll\kappa_\text{e}$. The dashed lines indicate $\kappa_{\rm ph}/\kappa_\text{e} = 2$ and $10$, which bound the region 
 of optimized thermoelectrics.}
\label{fig1}
\end{figure}

\noindent Fig.~\ref{fig1} visualizes the importance of the Wiedemann-Franz (WF) law for the past two centuries of thermoelectric research and condensed-matter science in general. Clearly, for the vast majority of experimentally documented materials, the WF law defines the upper bound of the TE figure of merit $zT$ for a given Seebeck coefficient $S$.
Here, we compiled more than 46,000 transport data entries from the StarryData2 database~\cite{katsura_starrydata_2025} covering the temperature range from 100 to 1000 K, and plot in Fig.~\ref{fig1} their $zT$ as a function of $S$. In the regime, where $\kappa_{\rm ph}\ll\kappa_{\rm e}$, the figure of merit simplifies to
\begin{equation}
    zT =\frac{S^{2}\sigma T}{\kappa_{\rm e}}\frac{1}{1+\kappa_{\rm ph}/\kappa_{\rm e}} \simeq \frac{S^{2}}{L}.
    \label{eq:zT_metal}
\end{equation}
For a conventional metal with $L=L_0$, this results in a direct correspondence between Seebeck coefficient and $zT$, which we define as the \textit{Wiedemann--Franz limit} shown as the black solid line in Fig.~\ref{fig1}. The WF limit can easily be extended within a single-parabolic band framework to account for nondegeneracy by using a well-known approximation \cite{kim_characterization_2015} $L=1.5+\exp[-|S|/116]$ instead of $L=L_0$, where $|S|$ is the absolute Seebeck coefficient in units of $\mu$V\,K$^{-1}$. Most high-performance thermoelectrics fall well below the WF limit because $\kappa_\text{ph}$ dominates over $\kappa_\text{e}$ in doped semiconductors due to their low -- compared to metals -- electrical conductivity. Conventional metals, on the other hand, approach the WF limit as even sizeable $\kappa_{\rm ph}$ becomes negligible compared to their large $\kappa_\text{e}$. Having said that, common metals often feature small Seebeck coefficients; only very recently,  $zT\rightarrow 1$ came into reach with the advent of \textit{metallic thermoelectrics} (METHEL) ~\cite{Garmroudi2023,garmroudi_topological_2025,garmroudi_energy_2025,Iwasaki2026}. In order to overcome the current limitations of METHEL, and thermoelectrics in general, novel strategies to reduce $L$ significantly below $L_0$ 
are required.

At first sight, the WF law appears to imply a fundamental trade-off between charge and heat transport, since increasing the electrical conductivity inevitably increases the electronic thermal conductivity. On the other hand, tuning $zT$ in metals by $\kappa_{ph}$ is \textit{ not necessary} at all, which simplifies the multi-parameter optimization problem of tuning $S$, $\sigma$ and $\kappa_{ph}$ in semiconductors to the necessity of enhancing $only$ the Seebeck coefficient in metals. The possibility of boosting $zT$ beyond the WF limit $S^2/L_0$ 
therefore depends on whether the Lorenz number can be tuned separately. Within linear response theory, all electronic transport coefficients, including the Lorenz number, are governed by a single energy-dependent transport distribution function $\Sigma(E)$~\cite{Ashcroft}. The transport coefficients are expressed through the energy integrals,
\begin{equation}
    \mathcal{L}_{n} = \int \Sigma(E)\,(E-\mu)^{n}
    \left(-\frac{\partial f_{0}}{\partial E}\right)dE,
    \label{eq:Ln}
\end{equation}
where $f_{0}(E)$ is the Fermi--Dirac distribution and $\mu$ is the chemical potential. In this notation, the electrical conductivity, Seebeck coefficient, and electronic thermal conductivity are given by

\begin{align}
    \sigma &= e^{2}\mathcal{L}_{0},\\
    S & =-\frac{\mathcal{L}_1}{eT\mathcal{L}_0},\\
    \kappa_{\rm e} &= \frac{\mathcal{L}_{2}}{T}-\sigma S^{2}T.
\end{align}

\noindent Using the WF law, the Lorenz number then follows as
\begin{equation}
    L=\frac{1}{e^{2}T^{2}}
    \left[
    \frac{\mathcal{L}_{2}}{\mathcal{L}_{0}}
    -
    \left(
    \frac{\mathcal{L}_{1}}{\mathcal{L}_{0}}
    \right)^{2}
    \right]
    =
    \left(
    \frac{k_{B}}{e}
    \right)^{2}
    \mathrm{Var}_{w}
    \left(
    \frac{E-\mu}{k_{B}T}
    \right),
    \label{eq:L_variance}
\end{equation}

\noindent where $\mathrm{Var}_{w}$ represents the transport-weighted variance of the reduced carrier energy (see SI for details). The Lorenz number therefore measures the energy spread of the carriers participating in charge and heat transport. Carriers farther from the chemical potential carry more heat per unit charge and contribute disproportionately more to $\kappa_{\rm e}$. Consequently, reducing the Lorenz number requires narrowing the energy distribution of the conducting carriers around the Fermi level $E_\text{F}$ rather than simply reducing their number.
This identifies the energy dependence of the transport distribution function as a crucial parameter governing the Lorenz number, since it determines how efficiently carriers at different energies contribute to charge and heat transport.

In semiclassical Boltzmann theory, the transport distribution function is determined by the electronic band structure and the dominant scattering mechanisms. In common metals and degenerate semiconductors, $\Sigma(E)$ varies only weakly over the thermal energy scale around $E_\text{F}$~\cite{Ashcroft, mckinney_search_2017}, and in the Sommerfeld expansion, one obtains the well-known Sommerfeld value
\begin{equation}
    L_{0} = \left(\frac{k_{B}}{e}\right)^2\!\left(\frac{\pi^2}{3}\right) \approx 2.44\times10^{-8} \ {\rm W\,\Omega\,K^{-2}}.
    \label{eq:L0}
\end{equation}
In contrast, for a non-degenerate semiconductor with an energy-dependent relaxation time, $\tau(E)\propto E^r$, the transport-weighted energy variance is reduced, yielding
\begin{equation}
    L_{\rm ndg} = \left(\frac{k_{B}}{e}\right)^{\!2}\!\left(\nu+\frac{5}{2}\right)\approx 1.49\times10^{-8} \ {\rm W\,\Omega\,K^{-2}},
    \label{eq:Lndg}
\end{equation}
for acoustic-phonon scattering ($\nu=-1/2$), which is noticeably lower than the Sommerfeld value. Together, $L_0$ and $L_{\rm ndg}$ define the conventional range of Lorenz numbers expected for metals and semiconductors. Between these two limiting cases, the Lorenz number evolves continuously with the degree of carrier degeneracy and is often well approximated by the interpolation proposed by Kim \emph{et al.}~\cite{kim_characterization_2015}.
\section{Optimal transport distribution function}
\begin{figure*}[t]
 \centering
 \includegraphics[width=2.0\columnwidth]{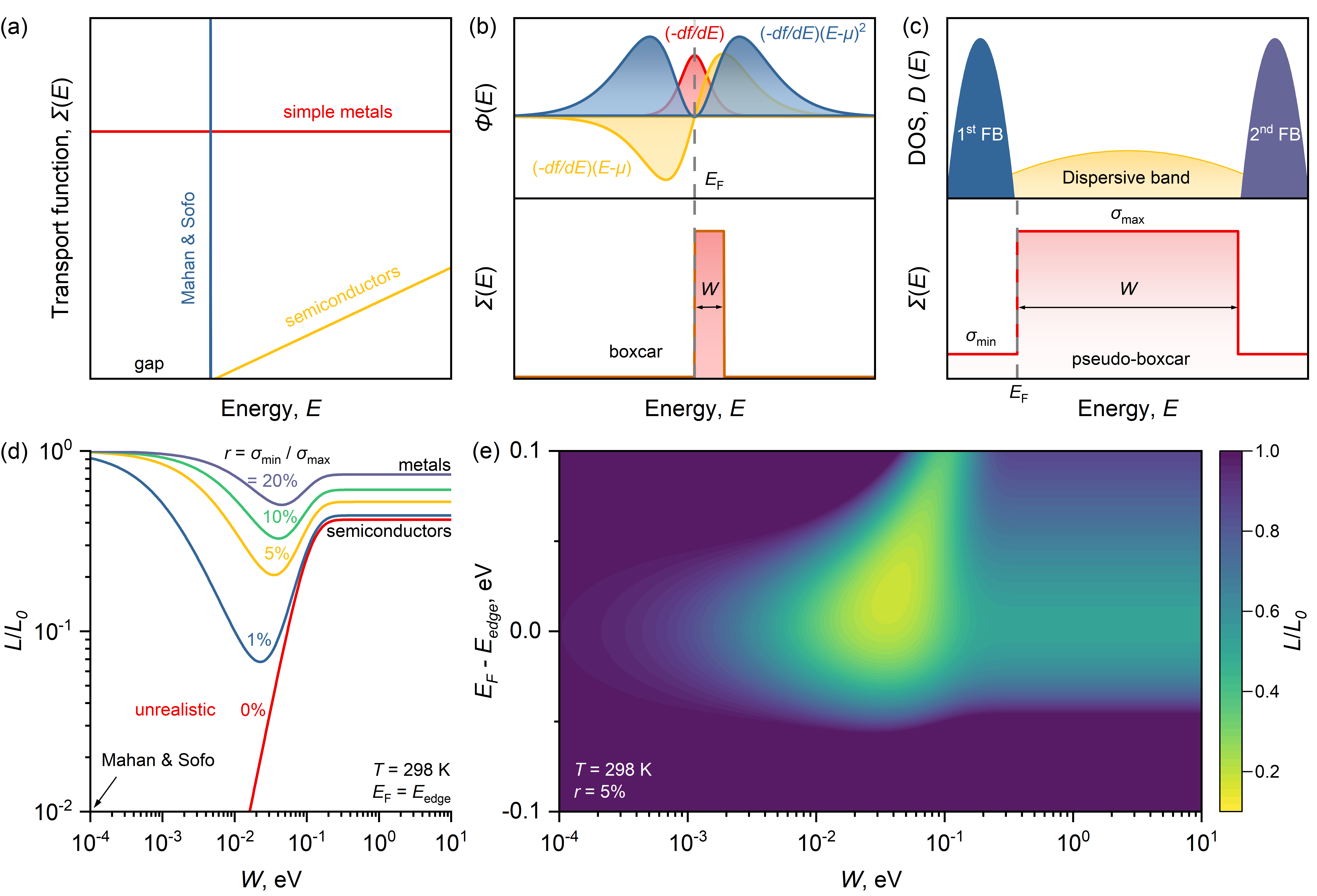}
 \caption{(a) Transport distributions for conventional metals, semiconductors, and the Mahan--Sofo limit~\cite{mahan_best_1996}. (b) Energy weighting of the transport coefficients for a boxcar transport distribution. (c) Sketch of a pseudo-boxcar transport distribution with flat-band-induced energy-selective scattering width $W$ and residual background $r=\sigma_{\min}/\sigma_{\max}$. (d) Calculated Lorenz number $L/L_0$ as a function of $W$ for different values of $r$. (e) False-color plot of the normalized Lorenz number $L/L_0$ as a function of the pseudo-boxcar width $W$ and edge position relative to the Fermi level, $E_{\rm F}-E_{\rm edge}$, at $r=5\%$ and $T=298$ K.
 }
\label{boxcar}
\end{figure*}

\noindent The optimal transport distribution for thermoelectric energy conversion has been the subject of theoretical investigation for nearly three decades. Within the idealized Boltzmann framework, Mahan and Sofo showed that the optimum corresponds to a delta-distribution-like $\Sigma(E)$ placed by a thermal energy scale away from $E_\text{F}$~\cite{mahan_best_1996}. In this limit, all carriers conduct at the same energy, the energy spread of conducting carriers vanishes, and the Lorenz number approaches zero at finite temperatures (see Fig.\ref{boxcar}a,d). Although this result establishes an important theoretical limit, a delta-distribution-like transport function cannot be realized in a real material because finite band dispersion and scattering inevitably broaden the conducting energy window. Also, a large pile-up of states usually gives rise to strong correlation effects, making a singular DOS unstable towards various types of electronic and magnetic ordering~\cite{tasaki}. Subsequent theoretical studies showed that imposing realistic constraints fundamentally changes the optimum transport distribution~\cite{fan_searching_2011,maassen_limits_2021,ding_best_2023}. Rather than a delta function, the highest thermoelectric performance is obtained for a finite-width boxcar transport distribution.

Although all transport coefficients are derived from the same function $\Sigma(E)$, they emphasize different energy regions (Fig.~\ref{boxcar}b). Electrical conductivity is determined primarily by carriers within the Fermi window $-\partial f_0/\partial E$, spaced over $\pm 2k_\text{B} T$ around $\mu$, whereas the Seebeck coefficient is additionally weighted by the displacement from the chemical potential through the factor $(E-\mu)$. In contrast, the electronic thermal conductivity weights carriers away from $\mu$ more strongly because of a quadratic $(E-\mu)^2$ term. As a result, carriers that contribute most to heat transport are not those responsible for most efficient electrical conduction. A boxcar transport distribution exploits this difference in energy weighting by restricting transport to a finite energy interval. It removes high-energy carriers that contribute disproportionately to $\kappa_{\rm e}$ while preserving those primarily responsible for electrical conduction. If the transport window is positioned asymmetrically with respect to the chemical potential, the resulting electron–hole asymmetry simultaneously generates a large Seebeck coefficient. Consequently, a boxcar-like $\Sigma(E)$ can reduce the Lorenz number without significant sacrifice in electrical conductivity or Seebeck coefficient~\cite{fan_searching_2011}.

\begin{figure*}[t]
    \centering
    \includegraphics[width=2.0\columnwidth]{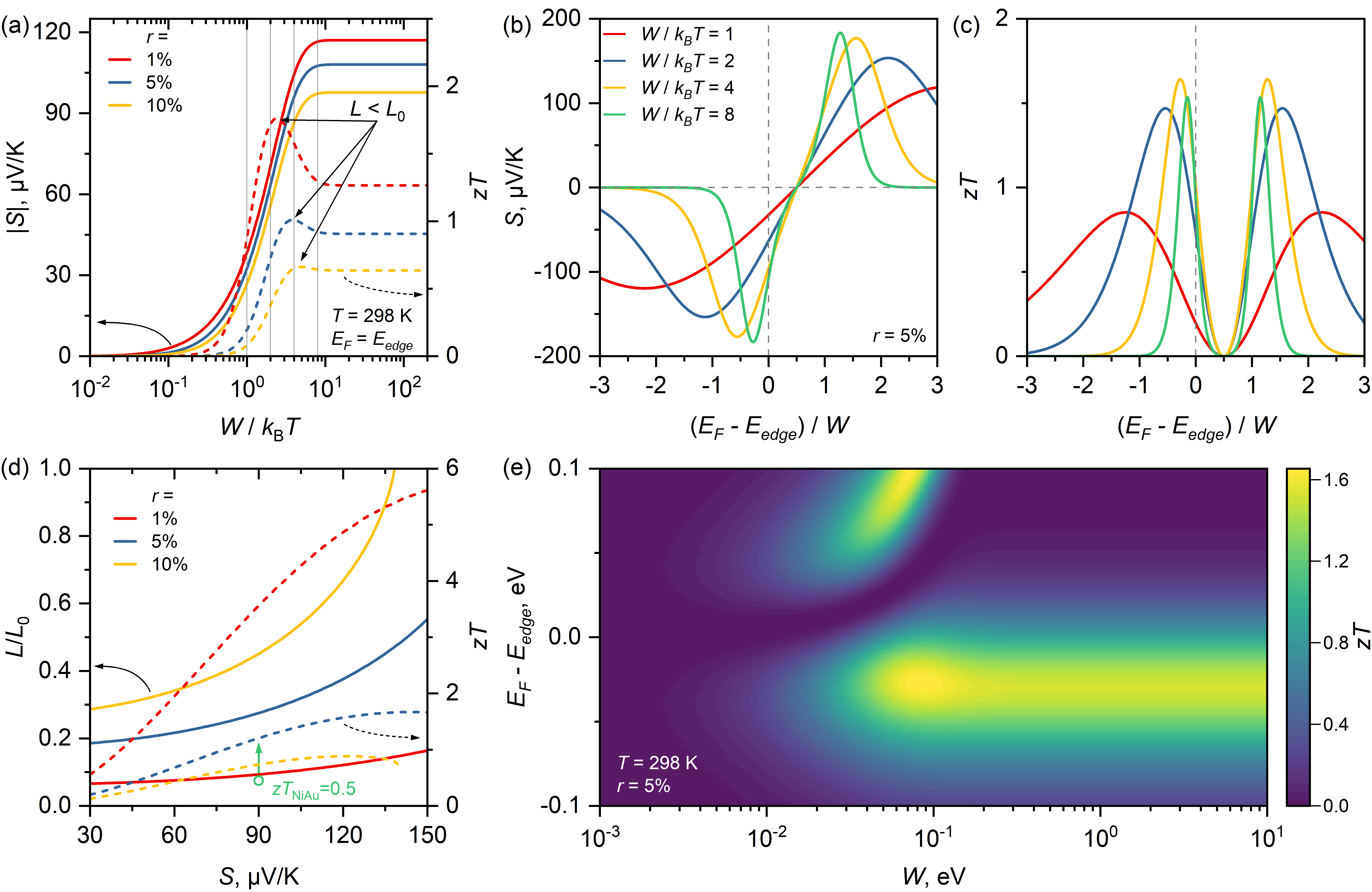}
    \caption{
(a) Evolution of $|S|$ (solid) and $zT$ (dashed) for a pseudo-boxcar transport distribution with scattering edge fixed at the Fermi level. (b,c) Dependence of $S$ and $zT$ on the position of the pseudo-boxcar relative to the chemical potential for different values of $W$ at $r=5\%$. (d) $L/L_0$ (solid) and $zT$ (dashed) versus $S$ for different $r$. The green circle marks the experimental $zT$ of Ni--Au alloys~\cite{Garmroudi2023}, and the arrow indicates the predicted 2.4-fold enhancement at fixed $S$ for $r=5\,\%$. (e) Heatmap of $zT$ as a function of the pseudo-boxcar width $W$ and edge position $E_{\rm F}-E_{\rm edge}$, at $r=5\%$ and $T=298$ K.
}
    \label{fig:S and zT}
\end{figure*}

\section{Pseudo boxcar transport via energy-separated flat bands}

\noindent The remaining challenge is therefore to identify a microscopic mechanism capable of generating a boxcar-like $\Sigma(E)$ in a real metal. We propose that energy-selective interband scattering of mobile carriers into less mobile flat-band states realizes this mechanism in a versatile, tunable fashion.

Flat bands are electronic bands with strongly reduced dispersion, which can arise via various mechanisms, from destructive quantum interference in frustrated lattice and orbital configurations \cite{kang2020topological,cualuguaru2022general,ye_hopping_2024,devarakonda2025frustrated}, $f$-electrons and the Kondo effect in heavy-fermion materials \cite{stewart1984heavy,fulde1988theory}, band folding in moir{\'e} heterostructures \cite{suarez_morell_flat_2010,li2021imaging,waters2024topological}, strong correlations in the presence of Hund's coupling \cite{oh2025hund}, or merely from localized, atomic-like wave functions \cite{regnault2022catalogue,garmroudi2022anderson,garmroudi2025energy,parzer2025mapping}. Their narrow bandwidth produces a large density of states confined to a small energy interval. At the same time, the strongly reduced group velocity of flat-band carriers makes their direct contribution to electrical transport small, yet they provide an energy-selective scattering channel for mobile carriers in dispersive conduction bands overlapping in energy with the flat-band states.

Let us consider first a simple metal in which transport is dominated by a single dispersive conduction band. In the absence of strong interband scattering, the relaxation time varies only weakly with energy, so that the transport profile remains nearly constant over the thermal transport window. Positioning a flat band on one side of the Fermi level opens an additional interband scattering channel once the flat-band states become energetically accessible. The resulting increase in scattering makes the relaxation time strongly energy dependent. Since $\Sigma(E)\propto D(E)v^{2}(E)\tau(E)$, the transport profile becomes sharply suppressed over the same energy interval. The nearly constant transport profile of a simple metal is thereby transformed into a step-like profile. When this suppression occurs only on one side of the Fermi level, either electron or hole transport is selectively reduced, naturally producing a large Seebeck coefficient~\cite{garmroudi_topological_2025,garmroudi_energy_2025}.

A second flat band separated by an energy $W$ introduces a complementary scattering edge. Between the two scattering edges, the dispersive band retains a high conductivity, whereas transport outside this interval is strongly suppressed by flat-band-induced scattering (Fig.~\ref{boxcar}c). The resulting profile is not an ideal boxcar, because the scattering edges are finite and the suppressed regions retain residual conductivity $\sigma_\text{min}$. We therefore refer to it as a \emph{pseudo-boxcar} transport profile, characterized by the width of the high-conductivity window, $W$, and the residual background $r=\sigma_{\min}/\sigma_{\max}$. The ideal boxcar is recovered only in the limit of perfectly sharp scattering edges and $r\rightarrow0$.

When $\mu$ is pinned to its low-energy edge, these two parameters completely determine the quality of the pseudo-boxcar transport profile and therefore its influence on $L$ and thermoelectric transport. Figure~\ref{boxcar}d summarizes the evolution of the Lorenz number as a function of the transport-window width $W$ for different residual backgrounds $r$ at $T = 298$~K. The Lorenz number approaches the Sommerfeld value in both the small- and large-$W$ limits, but for different reasons. For very narrow windows, the high-conductivity channel transmits only a small fraction of the thermally active carriers, so the current is dominated by the residual background. This background extends over a broad energy range and therefore retains the large transport-weighted energy variance characteristic of an ordinary metal. For sufficiently large $W$, the opposite limit is reached, where the conducting window spans most of the thermal transport range, energy filtering becomes ineffective, and the metallic transport-weighted energy variance is restored.

Between these two limits, the transport window is wide enough to carry most of the electrical current while remaining narrow enough to exclude carriers that dominate the electronic heat current (see Fig.~S3). This balance produces a pronounced minimum in the Lorenz number. The minimum occurs when the transport-window width is of the order of a few $k_{\rm B}T$: for example, for $T=298$~K, $r=1\%$, $L/L_0\simeq0.07$ at $W\simeq23 \text{ meV} \sim1 k_{\rm B}T$, whereas for $r=5\%$ the minimum shifts to $W\simeq38 \text{ meV} \sim 1.5$ $k_{\rm B}T$ with $L/L_0\simeq0.2$. Once the residual background exceeds approximately $10\%$, broad-band transport increasingly dominates, the transport-weighted energy variance is restored, and the Lorenz-number suppression becomes comparatively weak.

Can the Lorenz number be reduced even further? The analysis above assumes that one edge of the pseudo-boxcar is pinned near the Fermi level. The position of the conducting window relative to the chemical potential provides a second degree of freedom, as shown in Fig.~\ref{boxcar}d (see also Fig.~S2 for $r=1\%$ and $10\%$). Moving the pseudo-boxcar away from $E_\text{F}$ indeed produces an additional reduction of the transport-weighted energy variance. As the overlap between the conducting window and the Fermi window $-\partial f_0/\partial E$ decreases, transport becomes confined to a progressively smaller subset of thermally active carriers, and the calculated Lorenz number continues to decrease. The same displacement, however, simultaneously removes an increasing fraction of current-carrying states, causing $\sigma$ to decrease rapidly while $S$ increases. While, certainly, achieving $L\rightarrow 0$ is an intriguing scientific question~\cite{Stangier2022}, on its own it is therefore not the relevant optimization target for TE applications, as a large Seebeck coefficient is required at the same time. 

We therefore examine how the pseudo-boxcar parameters influence $S$ and the corresponding $zT$ as a function of the conducting-window width for different residual backgrounds (Fig.~\ref{fig:S and zT}). As $W$ increases, the second scattering edge progressively moves outside the thermal transport window, so the transport asymmetry becomes increasingly governed by the edge nearby the chemical potential. Consequently, the magnitude of the Seebeck coefficient increases rapidly before gradually approaching saturation.

The evolution of $zT$ is markedly different. Although $S$ continues to increase with $W$, the maximum $zT$ is reached at substantially smaller window widths. Increasing $W$ strengthens the transport asymmetry and enhances $S$, but progressively restores the transport-weighted energy variance. As a result, the reduction of the Lorenz number is partially lost. The optimum therefore occurs when the pseudo-boxcar remains a few $k_{\rm B}T$ wide, where $L$ is still strongly suppressed while $|S|$ has already reached most of its maximum value. Remarkably, increasing the residual background from 1\% to 10\% adversely affects both transport properties: it reduces $|S|$ and increases $L$, leading to a substantial decrease in $zT$. This establishes $r$ as a critical parameter for achieving high $zT$, because a finite background both restores the Lorenz number and degrades the transport asymmetry.

The role of $r$ as the second important parameter is illustrated in Figs.~\ref{fig:S and zT}b,c, where the pseudo-boxcar is translated relative to the chemical potential while its width remains fixed. When the transport profile is nearly electron-hole symmetric about $E_\text{F}$, the Seebeck coefficient vanishes. Shifting the pseudo-boxcar breaks this symmetry and produces either a positive or a negative $S$ depending on which scattering edge approaches the chemical potential. Consequently, $zT$ develops two maxima associated with the two scattering edges of the pseudo-boxcar. 

The promise of pseudo-boxcar transport lies in improving TE performance beyond what a large $S$ alone can deliver, as illustrated in Fig.~\ref{fig:S and zT}d. Ni--Au alloys provide a useful benchmark, reaching $zT\approx0.5$ with $S\approx-90\,\mu\mathrm{V/K}$\cite{Garmroudi2023}. Combining this already achievable $S$ with the suppression of $L$ enabled by a pseudo-boxcar profile could more than double $zT$, yielding $zT\approx1.2$ for $r=5\,\%$. Recent work on Ni$_3$In infers that a comparable value of $r$ can indeed be reached~\cite{garmroudi_topological_2025}, suggesting that the required scattering contrast may already be accessible, while identifying materials with still smaller $r$ could unlock substantially larger enhancements. Although this combination has yet to be realized in a single material, it would suggest that $zT\geq 1.5$ can be readily achieved via the Lorenz-number engineering presented here, offering a an exciting, largely unexplored route towards high-performance metallic TE.

\begin{figure*}[t]
 \centering
 \includegraphics[width=2.05\columnwidth]{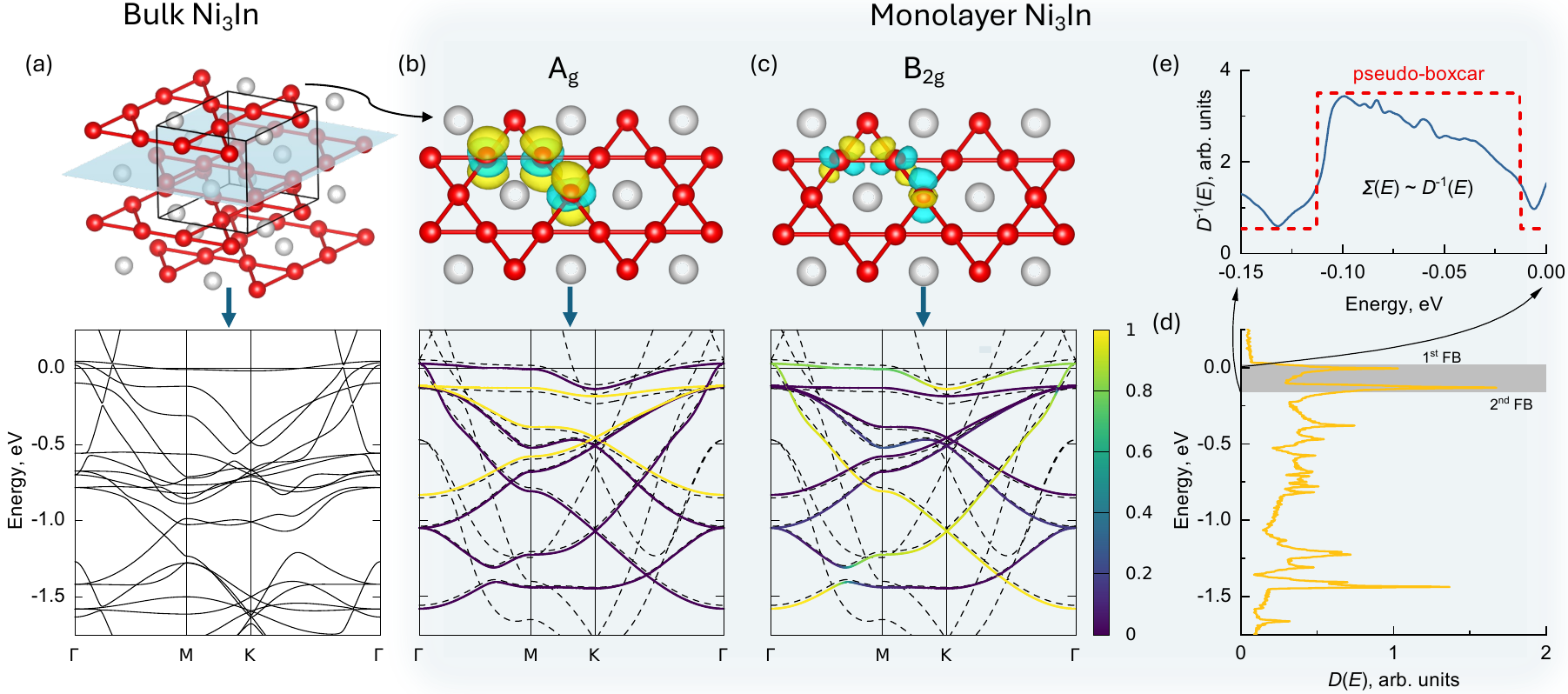}
\caption{Material realization of pseudo-boxcar transport in monolayer Ni$_3$In.
 Crystal structure and electronic band structure of (a) bulk and (b,c) monolayer Ni$_3$In. While bulk Ni$_3$In hosts a single kagome flat band formed by the in-plane Ni $d$ orbitals, removing the out-of-plane hybridization in the monolayer gives rise to an additional flat band derived from the Ni $d_z$ orbitals.
(d) Density of states of monolayer Ni$_3$In. The monolayer-induced $d_z$ flat band (2$^{\mathrm{nd}}$ FB) produces a sharp peak located approximately 0.12~eV below the in-plane flat band (1$^{\mathrm{st}}$ FB). (e) Shows the corresponding energy-dependent conductivity, $\Sigma(E)\propto D^{-1}(E)$, approaches a pseudo-boxcar transport profile when the Fermi level is positioned above the 1$^{\mathrm{st}}$ flat band.
}
\label{STF-resistivity-all}
\end{figure*}

Figure~\ref{fig:S and zT}e  summarizes the combined effect of width and position in a heat map, revealing a localized $zT$ maximum when the pseudo-boxcar is positioned slightly below $E_\text{F}$. The width and position therefore constitute two distinct design parameters: the width primarily controls the suppression of $L$, whereas the position determines the magnitude and sign of $S$. Together, these parameters define the design space for realizing pseudo-boxcar transport in metallic TE.

\section{Potential material platforms}

\noindent Among the wide variety of flat-band materials, the kagome metal Ni$_3$In provides a promising platform for realizing pseudo-boxcar transport. Bulk Ni$_3$In hosts the characteristic kagome band manifold, including a Dirac crossing, van Hove singularity, and a flat band formed by destructive interference of in-plane Ni $d$ orbitals~\cite{ye_hopping_2024}. Most importantly, the flat band in Ni$_3$In lies very close to the Fermi level and creates a sharply enhanced density of states over a narrow energy interval, providing the energy-selective scattering channel required to reshape the transport distribution~\cite{garmroudi_topological_2025}. In bulk Ni$_3$In, however, this mechanism supplies only one scattering edge albeit with an impressively low $r\approx 5\,\%$ ~\cite{garmroudi_topological_2025}. It can therefore generate a step-like transport profile and a large Seebeck coefficient, but not the finite-width pseudo-boxcar discussed above to reduce $L\ll L_0$.

To demonstrate that the proposed mechanism can be implemented in a real material, we performed first-principles calculations for monolayer Ni$_3$In. As shown in Fig.~\ref{STF-resistivity-all}, reducing Ni$_3$In to a single layer naturally generates the second flat band required for the proposed pseudo-boxcar mechanism. In bulk Ni$_3$In, the out-of-plane Ni $d_z$ states are dispersed by interlayer hybridization and do not form an isolated flat band. Removing this hybridization in the monolayer quenches the out-of-plane kinetic energy, producing an additional flat band predominantly derived from the Ni $d_z$ orbitals while preserving the original in-plane kagome flat band. The two flat-band onsets therefore define the boundaries of the high-conductivity transport window predicted by the model, and the calculated electronic structure naturally approaches the corresponding pseudo-boxcar transport profile. The calculated $d_z$ flat band is substantially narrower than the in-plane flat band and produces a sharper peak in the density of states approximately 0.12 eV below it, making the two-edge structure particularly distinct. Notably, the $d_z$ flat band is located below $E_\text{F}$, hence providing a natural source for a $p$-type METHEL -- a $desideratum$ on its own, as so far only $n$-type behavior could be realized~\cite{Iwasaki2026}. Electrostatic gating of a Ni$_3$In monolayer would then allow the chemical potential to be optimally positioned with respect to the two flat bands, providing a realistic route to tune the pseudo-boxcar window relative to $E_\text{F}$. Our first-principles calculations therefore demonstrate that the flat-band electronic structure of monolayer Ni$_3$In naturally realizes the transport design principles proposed above, providing a direct microscopic route toward Lorenz-number engineering in metallic thermoelectrics. They also hint at a more general design strategy, namely weakly coupled (van-der-Waals) kagome lattices, where flat bands due to destructive interference are accompanied by flat bands arising from the low-dimensional nature and quenched hopping across the layers.

\section{Discussion}
\noindent Here, we introduced a novel paradigm of reducing the Lorenz number in metallic TE by interband scattering of multiple flat bands, yielding a boxcar-shape transport distribution. While our initial inspection of Fig.~\ref{fig1} and its 3D visualization, Fig.S1, demonstrate the importance of the WF law, confining the vast majority of data points below the WF limit, we do find outliers that are 1--2 orders of magnitude above the WF limit. The crucial question is whether the outliers are measurement artifacts or $real$ violations of the WF law. An example for the former may be inhomogeneous materials, where determination of conductivity is ill-defined~\cite{Riss2024}, or anisotropic samples, e.g. unknowingly textured or strained polycrystalline materials, where $\sigma$ and $S$ are measured along the lateral direction of a pellet, whereas $\kappa$ is measured perpendicular to that~\cite{kim2015dense,deng_thermal_2018,chen2019enhancement}. A comprehensive case-by-case discussion of the data points above the WF limit in a future study may provide useful insights into possible new physical mechanisms that could severely reduce the Lorentz number. Even if all outliers turn out to be artifacts, our 'Wiedemann-Franz plot' of thermoelectrics displaying $zT$ vs $S^2$ [see also Fig.~1B in Ref.~\cite{Garmroudi2023}] provides a useful tool to check conformity of thermoelectric measurements. Scenario 1: if the data lie below the WF limit, everything appears consistent. Scenario 2: if the data lie above the WF limit, this indicates either an experimental inconsistency or unconventional transport physics. Regardless of its origin, scenario 2 requires a detailed reassessment of the experimental conditions, measurement geometry, sample homogeneity, etc.

In summary, we have established a general framework for reducing the Lorenz number in metallic thermoelectrics while preserving metallic electrical conductivity and a large Seebeck coefficient. A finite-width boxcar transport distribution has previously been identified as the theoretical optimum~\cite{ding_best_2023}, but here we show that its essential features can be reproduced through flat-band-induced energy-selective interband scattering. Two or more flat bands naturally generate the pair of scattering edges required for a pseudo-boxcar transport profile, whose width, residual background, and position determine the resulting thermoelectric performance. Finally, our first-principles calculations identify monolayer Ni$_3$In as a potential platform for realizing this concept, establishing flat-band engineering as a practical strategy for designing transport distributions in metallic thermoelectrics. We emphasize that this provides just $one$ possible route. Next to low-dimensional materials, separated flat bands can also be realized in three-dimensional materials, e.g. via crystal-electric-field splitting, magnetism, spin--orbit coupling or various other interactions lifting the degeneracy of a flat band.

\section{Methods}
\noindent The electronic structure of monolayer Ni$_3$In was investigated using first-principles density functional theory calculations as implemented in the Vienna ab initio simulation package (VASP), within the projector augmented-wave (PAW) method \cite{VASP,PAW1,PAW2}. The exchange-correlation interaction was described using the Perdew-Burke-Ernzerhof generalized-gradient approximation (PBE-GGA) \cite{PBE}. The Kohn-Sham wave functions were expanded in plane waves with an energy cutoff of 360\,eV, and the Brillouin zone was sampled using a $9\times9\times1$ Monkhorst-Pack $k$-point mesh. Electronic occupations were treated using a smearing width of 0.1\,eV. A vacuum region of 15\,\AA was introduced along the out-of-plane direction to avoid interactions between periodic images. The in-plane lattice constant was set to $a=5.06$\,\AA, and the pristine kagome geometry was retained without structural relaxation.

To construct an effective low-energy description, a nine-orbital Wannier model was obtained from the first-principles electronic structure \cite{wannier}. The basis comprises symmetry-adapted, rotated $d_{z^2}$, $d_{xz}$, and $d_{yz}$ orbitals, which form the $A_g$, $B_{2g}$, and $B_{3g}$ representations, respectively. The resulting Wannier Hamiltonian reproduces the relevant electronic bands in the vicinity of the Fermi level and was subsequently used to analyze the low-energy electronic structure.

\acknowledgments
\noindent I.\,S. acknowledges support by the Austrian Science Fund (FWF) 10.55776/ESP1731525.
F.\,G. acknowledges a Director’s Postdoctoral Fellowship through the Laboratory and Directed Research \& Development (LDRD) program. 
A.\,P. acknowledges support by the European Research Council (ERC Consolidator Grant 101231553- METHEL).

\bibliography{Literatur}
\bibliographystyle{apsrev4-2}

\end{document}